\documentclass[twocolumn,showpacs,preprintnumbers,amsmath,amssymb]{revtex4}

\usepackage{graphicx}
\usepackage{dcolumn}
\usepackage{bm}

\begin{document}

\preprint{APS/123-QED}

\title{Current and fluctuation in a two-state stochastic system under non-adiabatic periodic perturbation}
\author{Jun Ohkubo}
\email[Email address: ]{ohkubo@issp.u-tokyo.ac.jp}
\affiliation{
Institute for Solid State Physics, University of Tokyo, 
Kashiwanoha 5-1-5, Kashiwa-shi, Chiba 277-8581, Japan
}
\date{\today}

\begin{abstract}
We calculate a current and its fluctuation in a two-state stochastic system under a periodic perturbation.
The system could be interpreted as a channel on a cell surface or a single Michaelis-Menten catalyzing enzyme.
It has been shown that the periodic perturbation induces so-called pump current,
and the pump current and its fluctuation are calculated with the aid of the geometrical phase interpretation.
We give a simple calculation recipe for the statistics of the current, especially in a non-adiabatic case.
The calculation scheme is based on the non-adiabatic geometrical phase interpretation.
Using the Floquet theory, the total current and its fluctuation are calculated,
and it is revealed that the average of the current shows a stochastic-resonance-like behavior.
In contrast, the fluctuation of the current does not show such behavior.
\end{abstract}

\maketitle

\section{Introduction}

Recently, it has become possible to perform single-molecule experiments. \cite{English2006}
In such single-molecule experiments or small chemical systems such as cells,
it is expected that `fluctuation' plays an important role
because mean values and fluctuations (deviations) of observables are in the same order. \cite{Rao2002,Elowitz2002}
It also becomes possible to measure flux distributions experimentally in such small systems. \cite{Seitaridou2007}
When the fluctuation is large, the traditional rate equation approach is not adequate for the study of stochastic systems
because it treats only average values in large size limit.
Hence, the stochastic behavior of chemical reaction systems should be treated by a master equation 
approach. \cite{Risken1989,Gardiner2004}
Although it is difficult to obtain exact solutions of master equations in general,
it is important to evaluate (even approximately) not only the mean values, 
but also the fluctuations.

There is a phenomenon, known as a \textit{pumping},
in which a system under a periodic perturbation causes a finite flux in a preferred direction.
For example, a classical Michaelis-Menten (MM) enzymatic mechanism under a periodic perturbation
has been studied experimentally and theoretically.
It has been revealed that the periodic perturbation activates
a pumping mode in the MM system. \cite{Liu1990,Tsong2003,Astumian2003}
That is, the periodic perturbation causes a current
which is not explained by a simple average of those in the strict static cases.
The concept of the pump current is also related to molecular motors and Brownian ratchets. \cite{Julicher1997,Reimann2002}.

The average current can be calculated by various methods. \cite{Astumian2003,Astumian1989,Robertson1991,Jain2007}
In addition, it has been indicated that a calculation of the net current in the Brownian ratchets shows
a similar mathematical structure to that of the geometric phase in quantum mechanics. \cite{Astumian2002}
Recently, an explicit connection between the statistics of the current flow and the Berry phase interpretation
has been proposed. \cite{Sinitsyn2007, Sinitsyn2007a}
The total current is divided into the so-called `classical' current and `pump' current.
Sinitsyn and Nemenman have shown that the dynamical and the Berry phase 
correspond to the classical and pump current, respectively. \cite{Sinitsyn2007}
One of the other remarkable progresses is that 
it becomes possible to calculate not only the average current, but also the fluctuation about the current 
for an `adiabatic' cases.
Here, the `adiabatic' means that 
the oscillation of the periodic perturbation is very slow.

Although the average current has already been observed experimentally, \cite{Liu1990}
any experiment for the fluctuation has not been performed.
However, recent progress of experimental techniques \cite{Seitaridou2007} would make it possible to observe 
the flux distribution for the pumping phenomena directly.
Detailed information for non-equilibrium systems is useful to construct 
the non-equilibrium statistical mechanics and non-equilibrium thermodynamics. \cite{Ritort2008}
In order to investigate such non-equilibrium systems experimentally,
theoretical predictions would become a guideline for experimentalists.

In the present paper,
we give a calculation scheme for the statistics of the current,
especially in `non-adiabatic' cases.
The fluctuation about the average current in the non-adiabatic cases has not been calculated yet.
In addition, the non-adiabatic region is of great interest
because a stochastic resonance-like phenomenon has been observed experimentally 
in such non-adiabatic region. \cite{Astumian2003}
One of our main results is that there is no stochastic resonance-like behavior
for the fluctuation about the average current.
In order to calculate the statistics,
we develop a calculation scheme based on 
the non-adiabatic geometrical phase (Aharonov-Anandan phase \cite{Aharonov1987,Bohm2003}).
We have already proposed a method for direct evaluation of the Aharonov-Anandan phase
by means of a perturbation calculation for {\it eigenvectors} of a Floquet Hamiltonian. \cite{Ohkubo2008}
Here, we show that a perturbation calculation for {\it eigenvalues} of a Floquet Hamiltonian
gives the total statistics directly;
the calculation scheme is simpler than that in Ref.~21.

The present paper is organized as follows.
In Sec. II, we explain the stochastic model for the pumping phenomenon,
which has been introduced by Sinitsyn and Nemenman. \cite{Sinitsyn2007}
In Sec. III, we review a geometrical phase interpretation for the pumping phenomenon.
We develop an analytical treatment for the calculation of the total current
with the aid of the Floquet theory and the perturbation calculation in Sec. IV.
Section V gives concluding remarks.

\section{Stochastic pumps}

We mainly focus on a simple stochastic system with two absorbing states,
which has been introduced in Ref. 16.
The classical stochastic system is described as follows:
\begin{align}
[\mathrm{L}] \leftrightarrows [\mathrm{C}] \leftrightarrows [\mathrm{R}].
\end{align}
The system consists of three parts.
Two absorbing states are denoted as $[\mathrm{L}]$ and $[\mathrm{R}]$.
These absorbing states may be interpreted in many ways: 
they correspond to substrate and product in a MM enzymatic reaction, or cellular compartments, and so on.
These absorbing states exchange molecules or particles via an intermediate container $[\mathrm{C}]$.
The container $[\mathrm{C}]$ can contain either zero or one particle in it.
When the container is filled with one particle,
the particle can escape from the container by jumping into one of the two absorbing states, 
$[\mathrm{L}]$ or $[\mathrm{R}]$.
On the contrary, when the container is empty, either of the absorbing states
can emit a new particle into the container.
For simplicity, we here consider the following kinetic rates:
\begin{align}
\begin{array}{lll}
\textrm{(i)}   & [\mathrm{L}] \rightarrow [\mathrm{C}] : & k_1 = c_1 + R \cos (\omega t),\\
\textrm{(ii)}  & [\mathrm{L}] \leftarrow [\mathrm{C}] : & k_{-1} = c_{-1}, \\
\textrm{(iii)} & [\mathrm{C}] \rightarrow [\mathrm{R}] : & k_2 = c_2, \\
\textrm{(iv)}  & [\mathrm{C}] \leftarrow [\mathrm{R}] : & k_{-2} = c_{-2} + R \sin (\omega t), 
\end{array}
\end{align}
where $c_{\pm 1}$ and $c_{\pm 2}$ are positive real numbers, and $R$ is the amplitude of the perturbative oscillation.
The above kinetic rates indicate that 
only two kinetic rates ($k_1$ and $k_{-2}$) oscillate with time at a frequency $\omega$.

Note that the current $j$ at static cases ($\omega = 0$), in which all kinetic rates are constant,
is easily calculated from \cite{Sinitsyn2007}
\begin{align}
j = \frac{\kappa_{+} - \kappa_{-}}{K}, \quad K  \equiv \sum_{\{m\}} c_m, \quad \kappa_{\pm} \equiv c_{\pm 1} c_{\pm2}.
\label{eq_current_static}
\end{align}
We here describe the current $j$ as \textit{classical current}.
One might think that the time average of the current over a cyclic perturbation
is simply given by the time average of the classical current $j$, but it is not true.
For example, when $c_1 = c_{-1} = c_{2} = c_{-2}$ and $R \neq 0$, the time average of $j$ gives zero.
However, even in the simple case, a net current is actually observed.
Such additional current is called a \textit{pump current}.
The problem considered here is to calculate the \textit{total} current which includes both classical and pump currents.

\section{Phase Interpretation for the current statistics}

\subsection{Generating function for the current}

Our main goal is to calculate the net current between $[\mathrm{C}]$ and $[\mathrm{R}]$ in the steady state.
In order to calculate the current,
we use a method similar to the full counting statistics. \cite{Bagrets2006}
Let $P_n$ be the probability to have $n$ net transitions from $[\mathrm{C}]$ into $[\mathrm{R}]$ during time $T$,
where $T = 2\pi / \omega$ is the period of the rate oscillations.
The probabilities of a filled and empty state of $[\mathrm{C}]$ are denoted by $P_\mathrm{f}$ and $P_\mathrm{e}$ respectively,
and the state of the system is defined by
\begin{align}
\mathbf{p}(t) = 
\left[ \begin{array}{c}
P_\mathrm{e} \\ P_\mathrm{f}
\end{array}\right].
\end{align}
Due to the normalization condition, $P_\mathrm{e} + P_\mathrm{f} = 1$.
By discussions in Refs. 16 and 23,
the characteristic function of $P_n$ is given by 
\begin{align}
Z(\chi) &= e^{S(\chi)} = \sum_{s=-\infty}^{\infty} P_{n=s} e^{i s \chi} \notag \\
&= \mathbf{1}^\dagger \hat{T} \left( e^{-\int_{0}^{T} \hat{H} (\chi,t) d t}\right)
\mathbf{p}(0),
\label{eq_full_counting}
\end{align}
where $\mathbf{1}$ is the unit vector, $\hat{T}$ stands for the time-ordering operator, and
\begin{align}
\hat{H}(\chi,t) = 
\left[ \begin{array}{cc}
k_1 + k_{-2} & - k_{-1} - k_2 e^{i \chi} \\
-k_1 - k_{-2} e^{- i \chi} & k_{-1} + k_2
\end{array}\right].
\end{align}
In eq.~\eqref{eq_full_counting}, $\chi$ is called the counting field,
and the derivatives of $S(\chi)$ give cumulants of $P_n$,
e.g., $\langle n \rangle = - i \partial S(\chi) / \partial \chi |_{\chi=0}$.
Hence, the problem for calculating the current is actually the evaluation of the characteristic function $Z(\chi)$
(or $S(\chi)$).

\subsection{Interpretation as a Shr{\" o}dinger-like equation}

From eq.~\eqref{eq_full_counting},
it is easy to see that
the characteristic function $Z(\chi)$ is related to a solution of the following 
differential equation:
\begin{align}
\frac{d}{d t} \mathbf{p}(t) = - \hat{H}(\chi,t) \mathbf{p}(t),
\label{eq_time_evolution}
\end{align}
because the final state $\mathbf{p}(T)$ at time $T$ is formally calculated from
\begin{align}
\mathbf{p}(T)
= \hat{T} \left( e^{-\int_{0}^{T} \hat{H} (\chi,t) d t}\right) \mathbf{p}(0)
\equiv \exp[i \mu(\chi)] \mathbf{p}(0).
\end{align}
Using the normalization condition of the probability  $\mathbf{1}^\dagger \mathbf{p}(0) = 1$,
we obtain the identity $S(\chi) = i \mu(\chi)$.
Hence, the total current $J_\mathrm{total}$ and its derivative $J_\mathrm{total}^{(2)}$ are given by
\begin{align}
J_\mathrm{total} &= \frac{1}{T} \left. \frac{\partial \mu(\chi)}{\partial \chi} \right|_{\chi=0}, \\
J_\mathrm{total}^{(2)} &= (- i) \frac{1}{T} \left. \frac{\partial^2 \mu(\chi)}{\partial \chi^2} \right|_{\chi=0}. 
\end{align}
Although the existence of such cyclic state has not been mathematically justified in general cases,
we checked numerically that the stochastic system reaches a steady cyclic state, starting from arbitrary initial conditions. 
Note that when $\chi = 0$, eq.~\eqref{eq_time_evolution} simply gives a master equation for the time evolution of the system,
and hence 
\begin{align}
\mu(\chi = 0) = 0,
\label{eq_condition_mu}
\end{align}
due to the cyclic evolution of the system.

We here use an analogy between the classical stochastic system and a quantum mechanical formulation.
Replacing the time evolution operator $\hat{H}(\chi,t)$ by $H \equiv - i \hat{H}(\chi,t)$,
we obtain the following Shr{\" o}dinger-like equation: \cite{note_quantum}
\begin{align}
i \frac{d}{d t} | \phi (t) \rangle = H | \phi (t) \rangle
\label{eq_shrodinger}.
\end{align}
The time evolution operator $U(t)$ is constructed as
\begin{align}
| \phi(t) \rangle = U(t) | \phi(0) \rangle,
\end{align}
and the time evolution operator satisfies
\begin{align}
i \frac{d}{dt} U(t) = H U(t).
\end{align}
When we take the initial state $| \phi (0) \rangle$ as a cyclic state,
we have
\begin{align}
| \phi(T) \rangle = U(T) | \phi(0) \rangle = e^{i \mu(\chi)} | \phi(0) \rangle,
\label{eq_phase_introduction}
\end{align}
where $\mu(\chi)$ is a phase factor.
Hence, the state vector $| \phi(t) \rangle$ is an eigenstate of the time evolution operator $U(t)$,
and the phase factor $\mu(\chi)$ is related to its eigenvalue.

From this replacement,
the problem of calculation of the total current and its fluctuation
is replaced by the evaluation of the phase factor $\mu(\chi)$ for eq.~\eqref{eq_phase_introduction}.
Note that the `Hamiltonian' $H$ is a non-Hermitian operator,
which is different from the usual quantum mechanics.

\section{Direct evaluation of the total current and its fluctuation}

In order to calculate the total statistics, 
the Floquet theory is available. \cite{Moore1990,Moore1990a,Choutri2002}
The original Floquet theory for the geometrical phase was applied to Hermitian Hamiltonian cases, \cite{Moore1990,Moore1990a}
and it was extended to non-Hermitian cases. \cite{Choutri2002, Ohkubo2008}

As shown in Ref. 21, the `pump' current is calculated from the \textit{eigenvectors} of the Floquet states.
In contrast, we here shows that the `total' current is calculated from the \textit{eigenvalues} of the Floquet states directly.
The relationship among the total phase, the dynamical phase, and the Aharonov-Anandan phase
is shown in the Appendix.

\subsection{Usage of the Floquet theory}

From the Floquet theorem, 
the non-unitary time-evolution operator $U(t)$ for a periodic non-Hermitian Hamiltonian $H$ with period $T$
is decomposed into the Floquet product form as $U(t) = V(t) \exp(i M t)$.
Here, $U(t)$ is the unique fundamental matrix satisfying $U(0) = I$,
and the non-unitary matrix $V(t)$ also has a period $T$.
In addition, $M$ is a time-independent matrix.
By using the above Floquet product form, 
the total phase $\mu(\chi)$ is related to the eigenvalues of $U(T)$ via
\begin{align}
U(T) | \phi_\alpha(0) \rangle = e^{i M T} | \phi_\alpha(0) \rangle = e^{i \mu_\alpha (\chi)} | \phi_\alpha(0) \rangle,
\label{eq_Floquet_U}
\end{align}
where we define the initial state $| \phi_\alpha(0) \rangle$ as the eigenvector of $U(T)$.
The index $\alpha$ specifies an eigenvector and its corresponding eigenvalue.
Equation~\eqref{eq_Floquet_U} means that $\mu_\alpha(\chi) / T$ is the eigenvalues of $M$.
From the above discussion, 
it is needed to calculate the time evolution operator $U(t)$ or the matrix $M$
and then the evaluation of its eigenvalues should be done.
However, it is difficult to obtain them in general.

In order to evaluate the total phase $\mu_\alpha (\chi)$,
we here introduce another fundamental matrix and its Floquet product form $F(t) = P(t) \exp(i Q t)$,
in which $Q$ is assumed to be a diagonal matrix. \cite{Moore1990a,Ohkubo2008}
The fundamental matrix $F(t)$ is useful for the Fourier analysis.
Since $U$ and $F$ are both fundamental matrices of $H$,
there exists a constant invertible matrix $X$ with $U(t) = F(t) X$. \cite{Moore1990a}
Because $U(0) = I$, we have $U(0) = F(0) X = I$.
Hence, we obtain $X = F(0)^{-1}$ and 
\begin{align}
U(t) = F(t) F(0)^{-1} 
= P(t) P(0)^{-1} e^{i P(0) Q P(0)^{-1} t}.
\end{align}
Here, we can make identifications
$V(t) = P(t) P(0)^{-1}$ and $M = P(0) Q P(0)^{-1}$.
Using the above identifications and 
\begin{align}
M | \phi_\alpha (0) \rangle = \frac{\mu_\alpha (\chi)}{T} | \phi_\alpha (0) \rangle,
\end{align}
we obtain 
\begin{align}
Q |\alpha \rangle = \frac{\mu_\alpha (\chi)}{T} | \alpha \rangle,
\end{align}
where $| \alpha \rangle \equiv P(0)^{-1} | \phi_\alpha(0) \rangle$ is the eigenvector of $Q$.
Although the matrix $Q$ may not be taken as a diagonal matrix in some convenient bases in general cases,
in our case the matrix $Q$ can be diagonalized. \cite{Ohkubo2008}
Hence, the diagonal elements of $Q$ correspond to $\mu_\alpha (\chi) / T$ directly.
From the above discussions,
all we have to do is to calculate the eigenvalues of $Q$.

From the Floquet theory,
the matrix $Q$ is given by the eigenvalues of the Floquet Hamiltonian. \cite{Shirley1965}
The Floquet Hamiltonian $H_\mathrm{F}$ is defined by 
\begin{align}
\left\langle \alpha', n \right| H_\mathrm{F} \left| \beta, m \right\rangle
= H_{\alpha' \beta}^{(n-m)} + n \omega \delta_{\alpha'\beta} \delta_{nm},
\end{align}
where 
\begin{align}
H^{(n)} = \frac{1}{T} \int_0^T H e^{-i n \omega t} d t
\label{eq_fourier}.
\end{align}
The $| \alpha, n \rangle$ is an orthonormal basis for the matrix representation of $H_\mathrm{F}$.
The index $\alpha$ represents a `state part' and the index $n$ merely represents a Fourier component.
The `state part' means that the Hamiltonian is expressed as $2 \times 2$ matrix
and hence the eigenvector of the Hamiltonian has two components (states).
In our case, we denote the state parts as $+$ or $-$ (i.e., $\alpha, \beta \in \{ +, - \}$).
The $| \alpha, n \rangle$ is called as a ``Floquet state.''
Note that $H$ is non-Hermitian, so that $\langle \alpha', n |$ is not a conjugate state of $| \alpha, n \rangle$,
but the corresponding left orthonormal basis;
we add the prime on $\alpha$ in order to clarify the fact.

\subsection{Perturbative calculation for the eigenvalue}

Let $H_\mathrm{F}$ have eigenvectors $| \varepsilon_{\alpha,n} \rangle$ and eigenvalues $\varepsilon_{\alpha,n}$.
The matrix elements of $Q$ are therefore given by \cite{Shirley1965,Moore1990a}
\begin{align}
Q_{\alpha \beta} = - \varepsilon_{\alpha, 0} \delta_{\alpha \beta}.
\end{align}
In our case, the matrix form of the Floquet Hamiltonian $H_\mathrm{F}$ is written as follows:
\begin{widetext}
\begin{align}
\begin{bmatrix}
\ddots & \vdots & \vdots & \vdots & \vdots & \\
\ldots & -i c_1 - i c_{-2} + (n+1)\omega &  i c_{-1} + i c_2 e^{i \chi} & R(-i - 1)/2  &   0 & \ldots\\
\ldots & i c_1 + i c_{-2} e^{- i \chi} & - i c_{-1} - i c_2 + (n+1)\omega & R(i + e^{-i \chi})/2 & 0 & \ldots\\
\ldots & R(-i + 1)/2  &   0 & -i c_1 - i c_{-2} + n\omega  &  i c_{-1} + i c_2 e^{i \chi} & \ldots \\
\ldots & R(i - e^{-i \chi})/2 & 0 & i c_1 + i c_{-2} e^{- i \chi} & - i c_{-1} - i c_2 + n\omega & \ldots \\
 & \vdots & \vdots & \vdots & \vdots & \ddots\\
\end{bmatrix}
\end{align}
\end{widetext}
in the bases $\{\dots, |+, n+1 \rangle, |-, n+1 \rangle, |+,n\rangle, |-,n \rangle, \dots \}$.
It is easy to see that when $R = 0$, we have the block diagonalized matrix,
and the block diagonalized elements have a periodic behavior.
The block diagonalized Hamiltonian for $n = 0$ is given by
\begin{align}
H_\mathrm{F}^{(0)} = 
\begin{bmatrix}
-i c_1 - i c_{-2}     &  i c_{-1} + i c_2 e^{i \chi}  \\
 i c_1 + i c_{-2} e^{- i \chi} & - i c_{-1} - i c_2 
\end{bmatrix}.
\end{align}
The eigenvalues and right eigenvectors of the non-perturbative Hamiltonian $H_\mathrm{F}^{(0)}$
are denoted by $\varepsilon_{\alpha,0}^{(0)}$ and $| \varepsilon_{\alpha,0}^{(0)} \rangle$, respectively.
Since $H_\mathrm{F}$ is not Hermitian, we define the left eigenvectors of $H_\mathrm{F}^{(0)}$
as $\langle \varepsilon_{\alpha',0}^{(0)} |$.
From the eigenvalues of the block-diagonalized Hamiltonian $H_\mathrm{F}^{(0)}$,
we have the following total phase as the zeroth order of $R$:
\begin{align}
\mu_{\pm}^{(0)} (\chi) 
= i \frac{T}{2} \left[ K \mp \sqrt{K^2 + 4 (\kappa_{+} e_{+\chi} + \kappa_{-} e_{-\chi})} \right],
\label{eq_classical_phase}
\end{align}
where $e_{\pm \chi} \equiv e^{\pm i \chi} - 1$.
Equation~\eqref{eq_condition_mu} indicates that the index $+$ is adequate.
Hence, we select $\mu_{+}^{(0)} (\chi)$ as the total phase.
It is easy to see that the total phase for $R=0$ case recovers the static current $j$ (eq.~\eqref{eq_current_static}) adequately.

For $R \ll 1$ case,
the usual perturbation theory can be used to calculate $\mu_{+} (\chi)$.
There is no first order correction for the eigenvalues $\varepsilon_{+,0}$,
and hence we consider the second order correction:
\begin{align}
\varepsilon_{+,0}^{(2)} (\chi)= \sum_{\alpha \in \{+,- \} } \sum_{n\neq 0} 
\frac{
\langle \varepsilon_{\alpha',n}^{(0)} | H_\mathrm{F}' | \varepsilon_{+,0}^{(0)} \rangle
\langle \varepsilon_{+',0}^{(0)} | H_\mathrm{F}' | \varepsilon_{\alpha,n}^{(0)} \rangle
}
{\varepsilon_{+,0}^{(0)} - \varepsilon_{\alpha,n}^{(0)}},
\end{align}
where $H_\mathrm{F}'$ is the rest of the Floquet Hamiltonian $H_\mathrm{F}$,
which includes the perturbative parameter $R$.
After the perturbation calculation,
it is possible to calculate the total current $J_\mathrm{total}$ and 
its fluctuation $J_\mathrm{total}^{(2)}$ as follows:
\begin{align}
J_\textrm{total} &= - \frac{\partial}{\partial \chi} 
\left( \epsilon_{+,0}^{(0)}(\chi) + \epsilon_{+,0}^{(2)} (\chi) \right), \\
J_\textrm{total}^{(2)} &= i \frac{\partial^2}{\partial \chi^2}
\left( \epsilon_{+,0}^{(0)}(\chi) + \epsilon_{+,0}^{(2)} (\chi) \right).
\end{align}

\subsection{Results and discussions}

We here consider a simple case with $c_{1} = c_{-1} = c_{2} =c_{-2} = 1$.
In this case, the total current and its fluctuation up to the second order of $R$ are explicitly given as
\begin{align}
J_\mathrm{total} &= \frac{R^2}{4} \frac{\omega}{16 + \omega^2}, \label{eq_current_result} \\
J_\mathrm{total}^{(2)} &= \frac{1}{2} - \frac{R^2}{16 + \omega^2}. \label{eq_fluctuation_result}
\end{align}
From these expressions, it is easy to see the effects of the perturbation, as follows.

\begin{figure}
\begin{center}
  \includegraphics[width=80mm,keepaspectratio,clip]{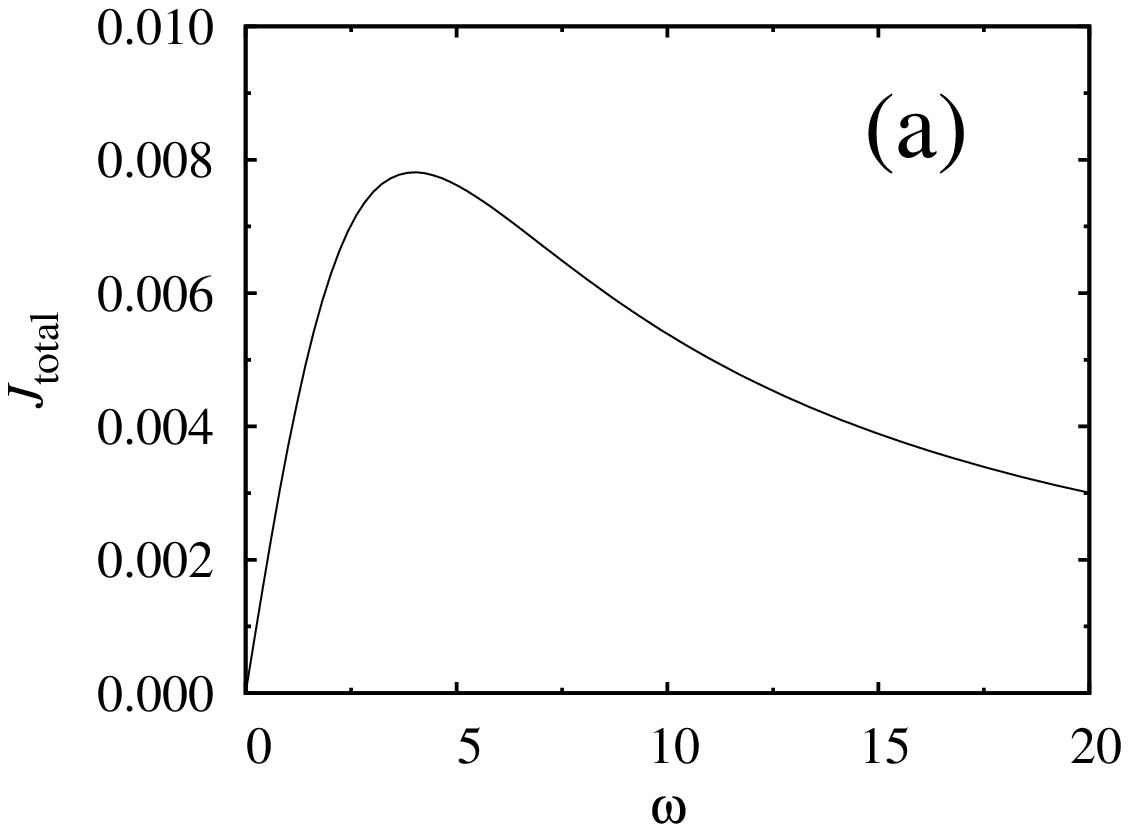} \\
  \includegraphics[width=80mm,keepaspectratio,clip]{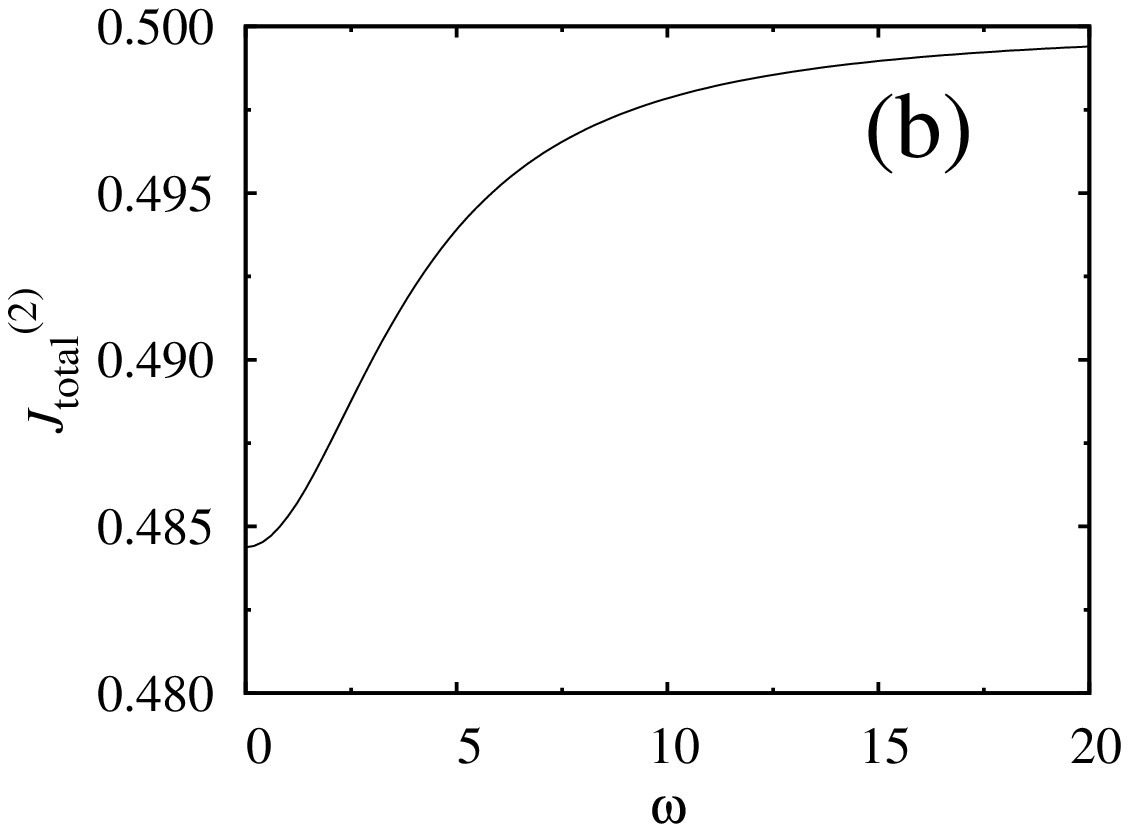} 
\caption{
(a) Total current $J_\mathrm{total}$ (eq.~\eqref{eq_current_result}).
(b) Its fluctuation $J_\mathrm{total}^{(2)}$ (eq.~\eqref{eq_fluctuation_result}).
Here, we set $c_1 = c_{-1} = c_2 = c_{-2} = 1$, and $R = 0.5$.
}
\label{fig_result}
\end{center}
\end{figure}

First, the perturbation induces a phenomenon like a stochastic resonance.
Figure \ref{fig_result}(a) shows the total current $J_\mathrm{total}$
calculated from eq.~\eqref{eq_current_result}.
Here, we set $R = 0.5$.
The total current shows a peak at a certain frequency $\omega_c$, 
which has been also observed experimentally. \cite{Liu1990,Astumian2003}
In the experiment in Ref.~7,
Na$^+$ pumping mode of (Na,K)-ATPase by an oscillating electric field has been studied.
Because different conformations of the protein have different dipole moments,
an oscillating electric field could drive structural change of the protein
and hence cause the modulation of kinetic parameters.
The actual dependency of the kinetic parameters on the external field could be very complicated,
and we assume that two of the kinetic rates depend on the oscillating field in our model.
Our model is a very simplified one,
so that the quantitative comparison with the experimental results may be difficult.
However, the qualitative behavior is the same as the experimental one:
there is one peak, and it decays as $\sim \omega^{-1}$ for $\omega \gg 1$. \cite{Astumian2003}
Hence, we believe that our model catches the feature of the pumping phenomenon.

Second, the perturbation varies the fluctuation of the total current, too.
While it is possible to evaluate the total current by various analytical methods, \cite{Astumian2003,Jain2007}
to our knowledge, the fluctuation has not been calculated explicitly for a long time.
In adiabatic cases, the fluctuation has been calculated via the Berry phase interpretation. \cite{Sinitsyn2007}
In the adiabatic cases, it has been revealed that the perturbation decreases the fluctuation of the current.
From eq.~\eqref{eq_fluctuation_result},
it is clear that the perturbation decreases the fluctuation of the total current even in the non-adiabatic case.
The fluctuation $J_\mathrm{total}^{(2)}$ is shown in Fig.~\ref{fig_result}(b).
In contrast to the current,
Fig.~\ref{fig_result}(b) shows no peak:
the fluctuation increases monotonically with the frequency $\omega$,
and approaches to the value of the non-perturbative case (in this case, $J_\mathrm{total}^{(2)} \to 1/2$).

Finally, we comment on the relationship between the present calculation and the Aharonov-Anandan phase.
Because the second order correction vanishes when we set $R=0$,
one might consider that all the second order correction stems from the Aharonov-Anandan phase.
However, it is not true.
For the simplest case with $c_1 = c_{-1} = c_2 = c_{-2} = 1$,
the classical current is zero, 
and the non-zero total current of eq.~\eqref{eq_current_result} stems from the Aharonov-Anandan phase.
In contrast, the fluctuation calculated from the Aharonov-Anandan phase becomes zero,
so that the second order term in eq.~\eqref{eq_fluctuation_result} is induced from the dynamical phase.
In summary, the second order correction does not stem from only the Aharonov-Anandan phase,
but also the dynamical phase.
If one wants to know only the `pumping' current and its fluctuation,
one more calculation for the Aharonov-Anandan phase is needed. \cite{Ohkubo2008}

\section{Conclusions}

In this work,
we gave a unified theory for the calculation of the total current and its fluctuation in classical stochastic systems
under a periodic perturbation of the kinetic rates.
The formulation is based on the full counting statistics and the non-adiabatic geometrical phase.
It was clarified that the total current is easily obtained by the combination of 
the Floquet theory and a simple perturbation calculation.
Although it is possible to calculate the Aharonov-Anandan phase directly, \cite{Ohkubo2008}
the calculation is a little complicated.
In a practical sense, the total current would be important,
and as shown in this work, the total current is easily obtained 
only by the perturbation calculation for the eigenvalues of the Floquet Hamiltonian;
the calculation is easier than the direct evaluation of the Aharonov-Anandan phase.
Hence, the formulation given in the present paper gives a useful analytical method
for the calculation of the current in the non-adiabatic cases.

Since our model is a simplified one,
we can obtain the analytical expressions for the total current and its fluctuation.
Although it may be difficult to compare the simple model and real experiments,
an important theoretical prediction is that there is no peak structure for the current fluctuation,
which is different from the behavior of the average current.
The simple model can predict the qualitative behavior for the average current
(one peak and the decay $\sim \omega^{-1}$) adequately,
so that we expect that the qualitative behavior for the fluctuation of the average current
obtained in the present paper will be observed in real experiments.
In addition, we hope that our analytical treatments and results 
become a basis for further studies,
especially for the connection with the non-equilibrium thermodynamics.

\section*{ACKNOWLEDGMENTS}
We thank N. A. Sinitsyn for useful comments,
and anonymous referees for many valuable information and suggestions.


\appendix
\section{Further discussions for relationship between dynamical phase and Aharonov-Anandan phase}

It is possible to divide the total phase $\mu(\chi)$ in eq.~\eqref{eq_phase_introduction} into two parts as follows: \cite{Bohm2003}
\begin{align}
\mu (\chi) = \delta (\chi) + \gamma (\chi).
\end{align}
$\delta(\chi)$ is called the dynamical phase, and $\gamma(\chi)$ is the Aharonov-Anandan phase.
The dynamical phase is defined by \cite{Bohm2003}
\begin{align}
\delta (\chi) = - \int_0^T \langle \widetilde{\phi} (t) | H | \phi (t) \rangle dt, 
\end{align}
where $| \phi(t) \rangle$ is a state vector of a cyclic evolved state, 
and $\langle \widetilde{\phi} (t) |$ is related to a evolved state associated with the periodic adjoint Hamiltonian $H^\dagger$.
The Aharonov-Anandan phase is calculated from \cite{Bohm2003,Choutri2002}
\begin{align}
\gamma (\chi) = \int_0^T \langle \widetilde{\phi} (t) | \frac{d}{dt}
| \phi (t) \rangle d t.
\label{eq_aharonov_anandan}
\end{align}
It has been shown that the Aharonov-Anandan phase gives the pump current. \cite{Ohkubo2008}
That is, there are the following correspondences between the phases and currents:
\begin{align}
J_\textrm{cl} \equiv \frac{1}{T} \left. \frac{ \partial \delta(\chi)}{\partial \chi} \right|_{\chi=0},\\
J_\textrm{pump} \equiv \frac{1}{T} \left. \frac{ \partial \gamma(\chi)}{\partial \chi} \right|_{\chi=0}.
\end{align}
The total current is given by $J_\mathrm{total} = J_\mathrm{cl} + J_\mathrm{pump}$.
Higher cumulants are also calculated from these phases.

In Ref.~21, we have developed a theory for the perturbative calculation for the eigenvectors of the Floquet state,
which is related to the state vector $|\phi(t) \rangle$.
In contrast, in the present paper,
we clarified that the total current is directly related to the eigenvalues of the Floquet Hamiltonian.
In practical sense, it is needed to evaluate the total current and its fluctuation,
because experimental observables include these two effects.
Although it is possible to evaluate the pump current from the Aharonov-Anandan phase, \cite{Ohkubo2008}
it is more convenient for us to calculate the total phase directly.


\begin{thebibliography}{99}
\bibitem{note_quantum}
One can continue the following discussions 
without introducing such analogy of the quantum mechanical formulation, by the use of the Fiber bundle.
Although the replacement of $\hat{H}$ by $H$ may have no physical meanings,
such quantum mechanical formulation has been developed well,
and hence we here use the replacement.
\end{thebibliography}
\end{document}